\documentclass[twocolumn,showpacs,superscriptaddress,10pt]{revtex4-1}
\usepackage{amsmath,amssymb,graphicx}
\usepackage{dcolumn}    
\usepackage{bm}         
\usepackage[caption=false]{subfig}
\begin{document}

\title{Enhanced superlens imaging with loss-compensating hyperbolic near-field spatial filter}

\author{Anindya Ghoshroy}
\author{Wyatt Adams}
\author{Xu Zhang}
\author{Durdu \"O. G\"uney}\email{Corresponding author: dguney@mtu.edu}
\affiliation{Department of Electrical and Computer Engineering , Michigan Technological University, 1400 Townsend Dr, Houghton, MI 49931-1295, USA}

\begin{abstract}
Recently a coherent optical process called plasmon injection ($\Pi$) scheme, which employs an auxiliary source, has been introduced as a new technique to compensate losses in metamaterials. In this work, a physical implementation of the $\Pi$ scheme on a thin silver film is proposed for enhanced superlens imaging. The efficacy of the scheme is illustrated by enhancing near-field imaging deeper beyond the diffraction limit in the presence of absorption losses and noise. The auxiliary source is constructed by a high-intensity illumination of the superlens integrated with a near-field spatial filter. The integrated system enables reconstruction of an object previously unresolvable with the superlens alone. This work elevates the viability of the $\Pi$ scheme as a strong candidate for loss compensation in near-field imaging systems without requiring non-linear effects or gain medium.
\end{abstract}
\maketitle

Controlling the interaction of light with metamaterials (MM) has instigated potentially revolutionary applications such as imaging beyond diffraction limit \cite{PhysRevLett.85.3966,zhang2008superlenses,Liu1686}, perfect absorbers \cite{aydin2011broadband}, electromagnetic cloaking \cite{pendry2009optics}, and many others. Realization of MMs has also advanced from microwave and lower frequencies to the visible spectrum \cite{valentine2008three} aided by technological advances in fabrication techniques. However, in the visible spectrum, MMs have significant dissipative losses in their constituent metallic structures which dampen the coupled oscillations of electrons and light (surface plasmon polaritons (SPPs)). For example, in the context of imaging beyond the diffraction limit, losses limit the resolution capabilities of "superlenses" hindering their advantages in applications such as nanoimaging and nanolithography \cite{kawata2009plasmonics}. Hence, overcoming losses is an important issue in MM research \cite{zheludev2010road}. Loss compensation schemes involving non-linear effects \cite{Popov:06} and electrically or optically pumped gain medium \cite{PhysRevLett.105.127401,xiao2010loss,Nezhad:04} were soon developed. Although these approaches have shown substantial promise from optical to terahertz frequencies, their usefulness has been limited in the context of imaging. This is because the preservation of amplitude and phase relationships between fields in time and space is difficult with gain-assisted MMs.

Our approach to loss compensation employs a secondary illumination or "auxiliary source." This source externally injects SPPs to amplify the local, lossy SPP eigenmodes of a plasmonic MM. The technique was initially conceptualized for a single wavevector in \cite{sadatgol2015plasmon}. An auxiliary source capable of amplifying an arbitrary wavevector was also envisioned and loss compensation scheme was named the "plasmon-injection" (PI) or $\Pi$ scheme. Theoretical studies later showed that the technique is similar to a linear deconvolution process \cite{Adams:17} capable of enhancing \cite{Adams:17,adams2016bringing,zhang2016enhancing,Zhang:17} the resolution limits of previously studied near-field imaging systems employing negative index MMs, plasmonic lenses and hyperlenses. However, in a physical implementation of the $\Pi$ scheme, a detailed understanding of the properties of the auxiliary source was necessary. These were later presented in \cite{Ghoshroy:17} where the word "active" is used to distinguish the $\Pi$ scheme from previous "passive" counterparts \cite{Adams:17,adams2016bringing,zhang2016enhancing,Zhang:17}. It was shown that the auxiliary must provide amplification to a narrow band of high spatial frequency components to avoid amplifying noise and must be constructed by a physical convolution with the object's field distribution. Following this revelation, it was shown how a hyperbolic metamaterial (HMM) functioning as a tunable near-field spatial filter presents a possible method to construct an auxiliary source which has the above mentioned properties \cite{ghoshroy2017hyperbolic}.

This letter describes for the first time, a physical implementation of the active $\Pi$ scheme. A HMM spatial filter is integrated into a near-field imaging system to construct the convolved auxiliary source with selective amplification capabilities and an iterative reconstruction process, previously discussed in \cite{Ghoshroy:17}, and is used to reconstruct the image spectrum. A plasmonic imaging system operating at wavelength $\lambda = 365 \ nm$ and employing a $50 \ nm$ thick silver lens is selected. This superlens is incapable of resolving subwavelength features at one-sixth of the illumination wavelength $(\lambda/6)$ primarily due to material losses and noise. Numerical calculations, performed with the finite element based software package COMSOL Multiphysics, show how the active $\Pi$ scheme when implemented with the system allows the resolution of subwavelength features separated by $\lambda/6$. Additionally, the reconstruction process requires no prior knowledge of the object and the concept can be extended to other near-field imaging systems at different wavelengths.

\begin{figure}[htbp]
\centering
\includegraphics[height=0.3\textwidth]{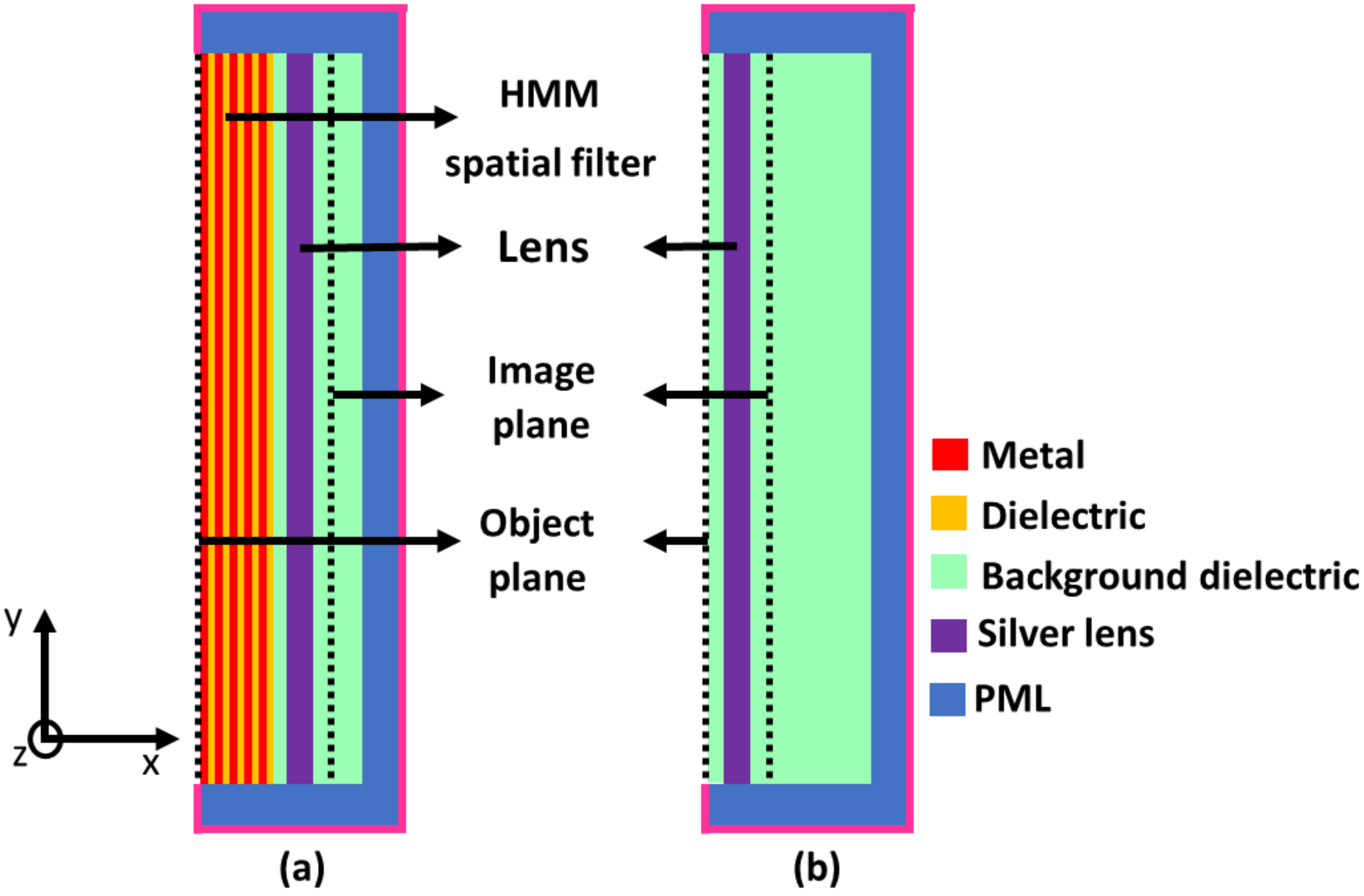}
\caption{The geometry of the imaging system, (a) with, and (b) without  the integrated spatial filter, built in COMSOL to perform numerical simulations (not to scale). The object is defined as a magnetic field polarized along the z-axis along the object plane. The response is extracted from the image plane. The red and yellow regions are the metallic and dielectric layers, respectively. The green, magenta and blue regions are the background dielectric, silver lens, and perfectly matched layer (PML), respectively. Scattering boundary conditions are applied to the edges of the PMLs and are highlighted in pink.}
\label{fig:Fig_Imaging_System_Schematic}
\end{figure}

A schematic of the silver lens imaging system, with and without the integrated HMM spatial filter is shown in figures \ref{fig:Fig_Imaging_System_Schematic}(a) and (b), respectively. The spatial filter is constructed by alternately stacking layers of aluminum with dielectrics. The imaging system is embedded inside a dielectric with relative permittivity $\epsilon _d = 2.5$ similar to an experimental silver lens \cite{fang2005sub}. The relative permittivities of silver and aluminum at $\lambda = 365 \ nm$ are $\epsilon _{Ag} = -1.8752 -i0.5947$ and $\epsilon _{Al} = -18.179 -i3.2075$, respectively calculated from the Drude-Lorentz model \cite{rakic1998optical}. The systems are excited with a transverse magnetic (TM) polarized field from the object plane and the responses are extracted as a complex magnetic field from the image plane. The object and image planes are marked by dashed lines in figures \ref{fig:Fig_Imaging_System_Schematic}(a) and (b). Both geometries are padded with perfectly matched layers (PMLs) shown in blue while the edges of the PMLs are backed by scattering boundary conditions shown by pink lines. The extent of the geometry along the y-axis is $80\lambda$ and $9500$ mesh elements are defined at each boundary parallel to the y-axis with the smallest mesh element being approximately equal to $3 \ nm$. Transfer functions are calculated by using a TM point source excitation from the object plane. A Gaussian field distribution with full width at half maximum (FWHM) of $6 \ nm$, negligibly small compared with the operating wavelength, is used to mimic the point source. The response of the system is the point spread function, the Fourier transform of which gives the complex transfer function. Prior to all image processing calculations, noise is introduced into the measured fields from the image plane by assuming that the spatial distribution of the field is measured by a detector consisting of an array of pixels. The signal recorded at each pixel is distorted by a combination of signal dependent (SD) and signal independent (SI) noise processes. The corresponding noisy image is calculated according to the ``signal-modulated noise" model \cite{Walkup,Heine:06,Froehlich:81}, employed previously in \cite{Ghoshroy:17}. This work uses $0.025I(y) \ A/m$ and $0.005 \ A/m $ for the standard deviations of SD and SI noise, respectively and $I(y)$ is the spatial distribution of the noiseless or ideal signal on the image plane. Note that these values are larger than the standards adopted in \cite{Chen:16,Ghoshroy:17} as well as an experimental optical detector \cite{Akiba:10} and are selected to highlight the adequacy of the active $\Pi$ scheme in noisy systems.

\begin{figure}[h]
\centering
\includegraphics[height=0.25\textwidth,width=\linewidth]{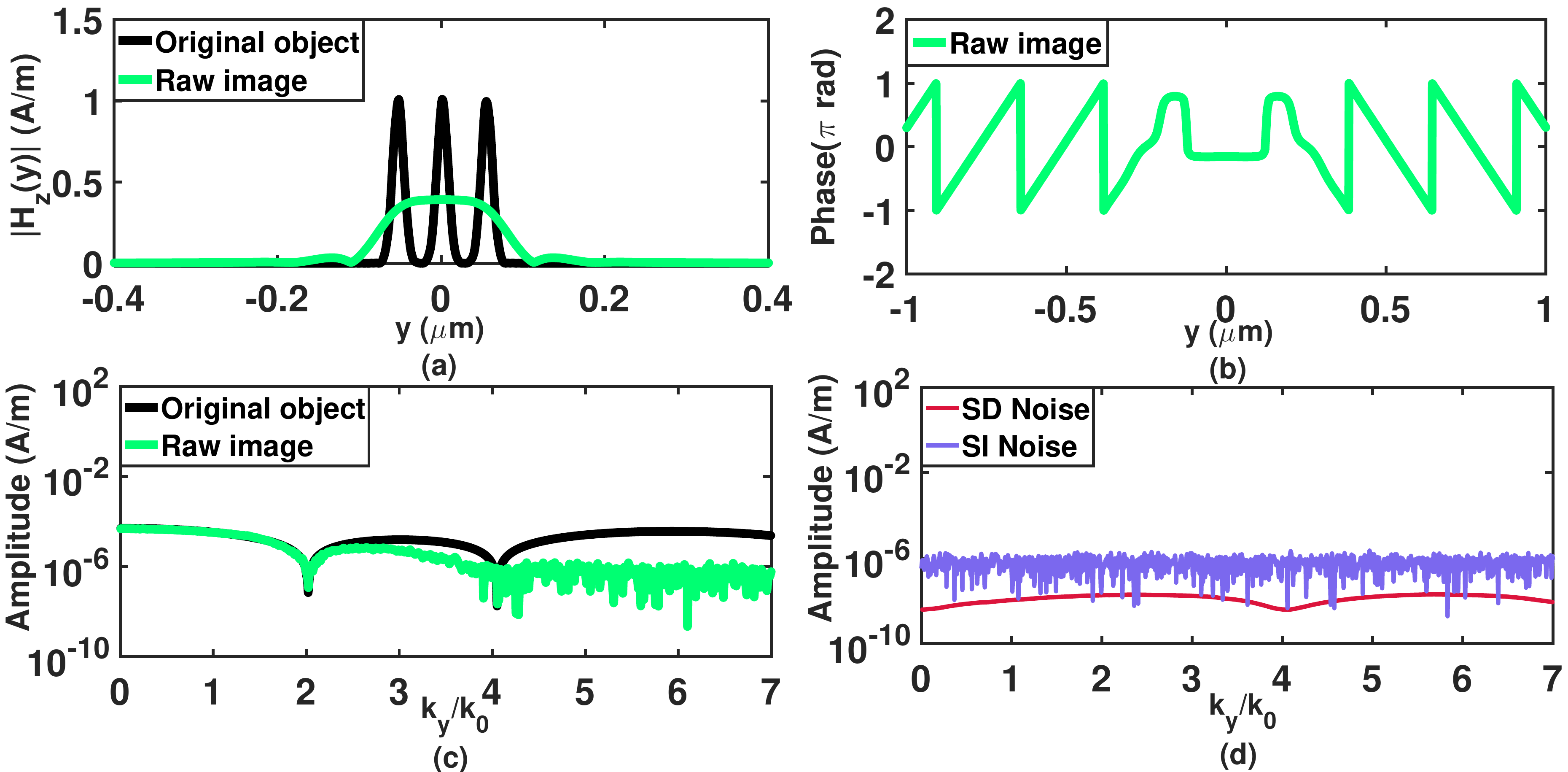}
\caption{(a) Amplitude of the magnetic field distribution of the object and the raw image from the object and image planes (see figure \ref{fig:Fig_Imaging_System_Schematic}(b)). (b) The phase of the magnetic field of the raw image. The object is purely real and has zero phase. (c) The amplitude of the Fourier transform of the object and the raw image showing how the losses and noise have corrupted the image spectrum beyond $k_y = 2k_0$. (d) The contributions of the SD and SI noise to the raw image spectrum.}
\label{fig:Fig_FFT_Field_Distributions_Initial}
\end{figure}

The usefulness of the active $\Pi$ scheme in an imaging scenario is exemplified with an arbitrary object with $3$ Gaussian features separated by $\lambda/6$. The object is defined as a real TM polarized field along the object plane (see figure \ref{fig:Fig_Imaging_System_Schematic}). In a physical imaging system, no prior knowledge of the object will be available. Therefore, only the raw data from the image plane is used to pinpoint the spatial frequency at which the first instance of the auxiliary source should be applied. For this purpose, the silver lens without the integrated spatial filter (see figure \ref{fig:Fig_Imaging_System_Schematic}(b)) is used to image the object field. The measured amplitude and phase of the raw image are corrupted with noise and are shown by the green lines in figures \ref{fig:Fig_FFT_Field_Distributions_Initial}(a) and (b), respectively. It is evident that the silver lens is incapable of resolving the object shown by the black line in figure \ref{fig:Fig_FFT_Field_Distributions_Initial}(a). The amplitude of the Fourier transforms of the object and the raw image are shown in figure \ref{fig:Fig_FFT_Field_Distributions_Initial}(c) while figure \ref{fig:Fig_FFT_Field_Distributions_Initial}(d) shows the contributions of the SD and SI noise which distort the raw image spectrum. Figure \ref{fig:Fig_FFT_Field_Distributions_Initial}(c) shows how the losses and noise start to progressively degrade the image spectrum from $k_y \approx 2k_0$, where $k_0$ is the free-space wavenumber. Eventually, the image spectrum is completely overwhelmed by the noise and cannot be recovered by passive deconvolution or increased illumination intensity since both processes proportionally amplify noise \cite{Ghoshroy:17}. In conclusion, a selective amplification with the auxiliary must be initiated from $k_y \approx 2k_0$ and progressively moved to higher spatial frequencies. Note that the Fourier spectra in figures \ref{fig:Fig_FFT_Field_Distributions_Initial}(c) and (d) are truncated at $k_y = 7k_0$ because the calculated transfer function starts to lose accuracy beyond this point. Hence, the selective amplification process can be applied till $k_y = 7k_0$. One important reason for the failure of the numerically calculated transfer function is the finite spatial extent of the image and object planes which introduces errors in the Fourier transform calculations as discussed in \cite{Ghoshroy:17}.

\begin{figure}[h]
\centering
\includegraphics[width=\linewidth]{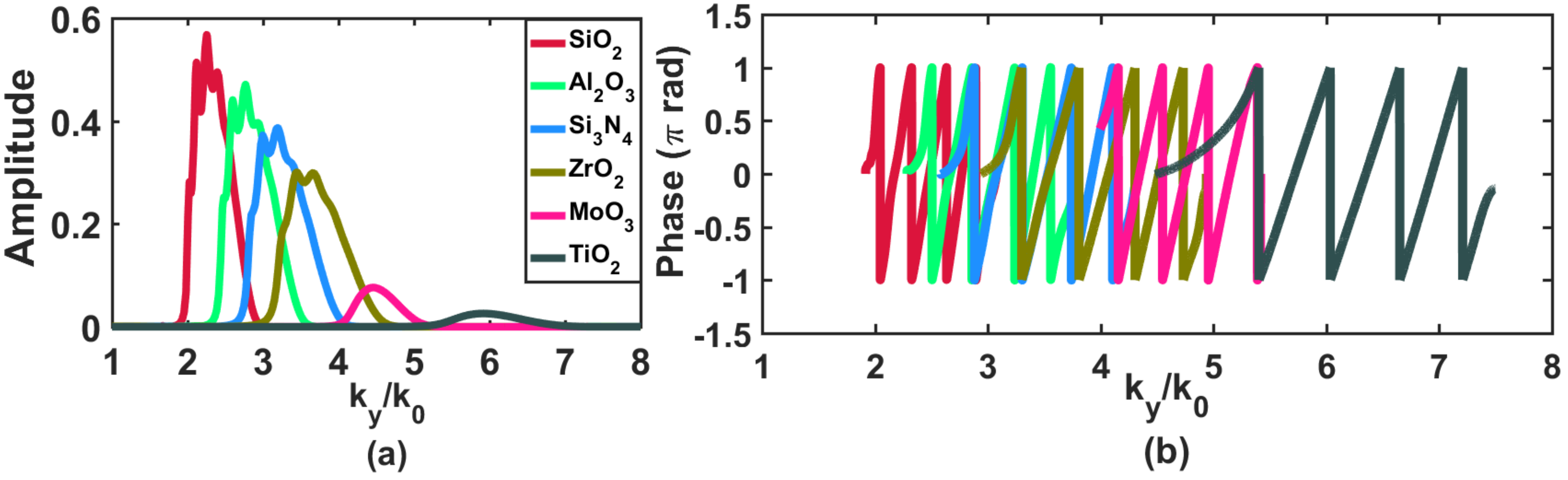}
\caption{(a) Amplitude and (b) phase of the transfer function of each HMM spatial filter. The phase plot shows only the region of high transmission since the calculated phase is irrelevant and not reliable outside the pass-band of the filter due to poor transmission.}
\label{fig:Fig_HMM}
\end{figure}

In order to design an auxiliary source to cover the spectral band $2k_0 \leq k_y \leq  7k_0$, tunable HMM spatial filters with pass-bands within this spectral range will be used below. As shown in \cite{ghoshroy2017hyperbolic}, the transmission band of the HMM spatial filter can be tuned to different spatial frequencies by changing the relative permittivity of the constituent dielectric layer and the filling fraction of each unit cell. Therefore, six HMM spatial filters are designed to cover the high spatial frequency range $2k_0 \leq k_y \leq  7k_0$. The dielectrics, their relative permittivities along with the filling fraction and number of unit cells of each HMM which are used to construct these spatial filters are shown in table \ref{tab:Table_Dielectrics}. The metallic layer is set to aluminum and the overall thickness of the HMM is kept constant at $365 \ nm$. The relative permittivities for the dielectrics listed in table \ref{tab:Table_Dielectrics} at $\lambda = 365 \ nm$ are obtained from \cite{Gao:12,Kelly:72,Luke:15,Wood:82,PhysRevB.88.115141,ADOM:ADOM201600250}. The amplitude and phase of the complex transfer function corresponding to the HMM spatial filters are shown in figures \ref{fig:Fig_HMM}(a) and (b), respectively. Note that only the phase within the pass-band of each spatial filter is shown in figure \ref{fig:Fig_HMM}(b) since the calculated phase outside this pass-band is irrelevant and inaccurate because of poor transmission (approximately five orders of magnitude less).
\begin{table}[h]
\small
\centering
\caption{\bf Design parameters of the HMM spatial filters.}
\begin{tabular}{cccc}
\hline
Dielectric & Relative     &   Number of  & Filling  \\
           & permittivity &   unit cells & fraction  \\
\hline
$SiO_2$     & $2.2147$             & $10$ & $0.65$ \\
$Al_2O3$    & $3.18587$            & $10$ & $0.65$ \\
$Si_3N_4$   & $4.05373$            & $10$ & $0.65$ \\
$ZrO_2$     & $5.06205$            & $10$ & $0.65$ \\
$MoO_3$     & $6.031 - 1.1908i$    & $10$ & $1.1$  \\
$TiO_2$     & $8.2886 - 0.10186i$  & $12$ & $1.1$  \\
\hline
\end{tabular}
\label{tab:Table_Dielectrics}
\end{table}

\begin{figure}[htbp]
\centering
\includegraphics[height=0.25\textwidth,width=\linewidth]{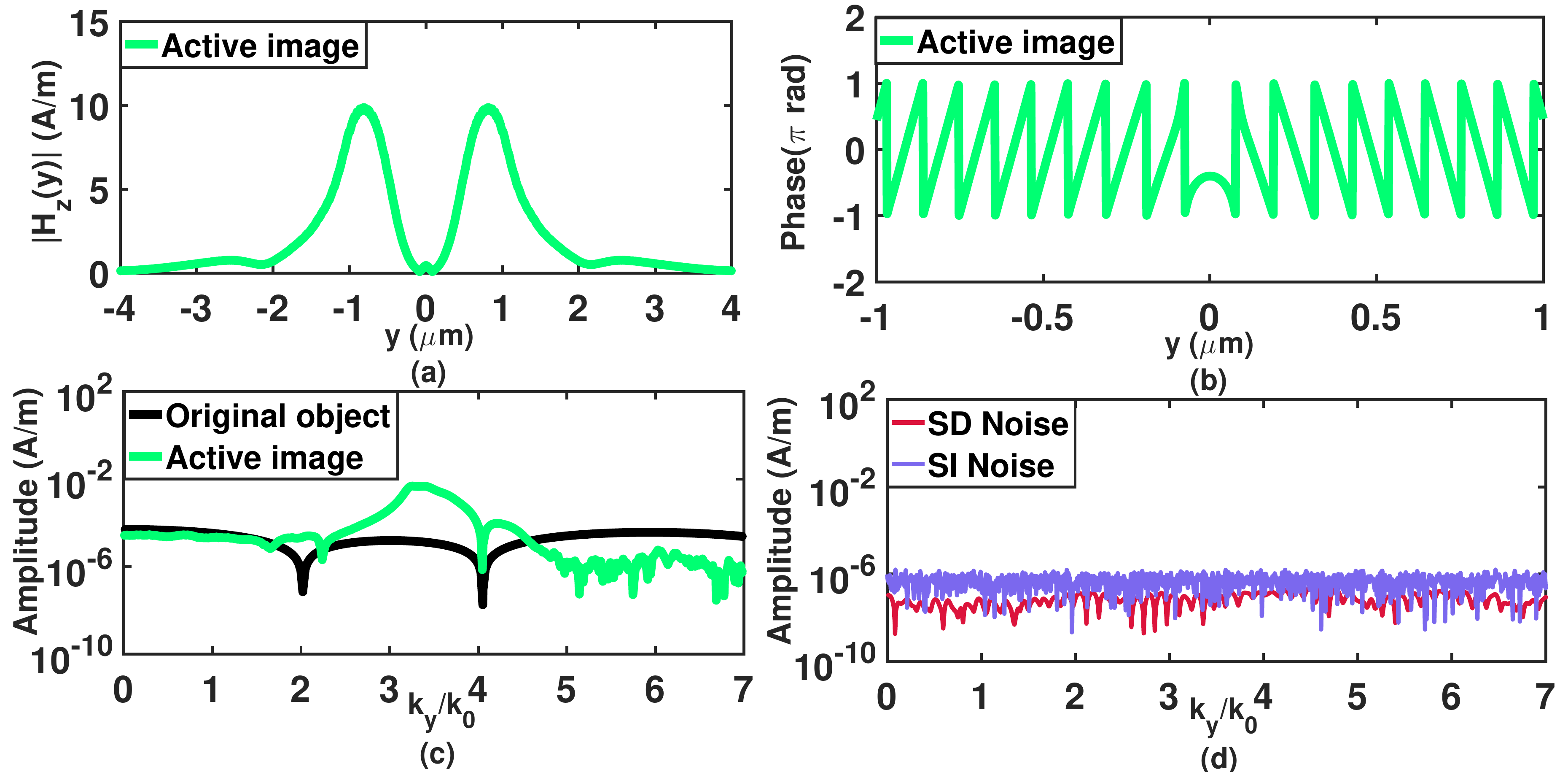}
\caption{(a) Amplitude and (b) phase of the magnetic field distribution of the active image from the image plane (see figure \ref{fig:Fig_Imaging_System_Schematic}(a)). (c) The amplitudes of the Fourier transform of the object and the active image show how the selective amplification has recovered a previously undetectable object feature at  $k_y = 4k_0$. (d) The contributions of the SD and SI noise spectra. The level of the SD noise is approximately the same as figure \ref{fig:Fig_FFT_Field_Distributions_Initial}(d).}
\label{fig:Fig_FFT_Field_Distributions_SiO2}
\end{figure}

It is instructive to illustrate the effect of selective amplification with an auxiliary source with one filter-lens system. Therefore, we select the integrated $Al-ZrO_2$ HMM spatial filter (see figure \ref{fig:Fig_HMM}). The object is illuminated with a coherent light source of sufficiently high intensity and then imaged with the integrated system (see figure \ref{fig:Fig_Imaging_System_Schematic}(a)). The corresponding image measured by the detector is called the \emph{"active image."} The amplitude and phase of the active image are shown in figures \ref{fig:Fig_FFT_Field_Distributions_SiO2}(a) and (b), respectively. The Fourier transform of the active image and the object spectrum are shown by green and black lines, respectively, in figure \ref{fig:Fig_FFT_Field_Distributions_SiO2}(c). Figure \ref{fig:Fig_FFT_Field_Distributions_SiO2}(d) shows the contributions of the SD and SI noise which distort the active image spectrum. Comparing the active image spectrum in figure \ref{fig:Fig_FFT_Field_Distributions_SiO2}(c) with the raw image spectrum in figure \ref{fig:Fig_FFT_Field_Distributions_Initial}(c), it is evident that the noise is not visible in the active image spectrum within the pass-band of the $Al-ZrO_2$ spatial filter whose transfer function is shown in figure \ref{fig:Fig_HMM}(a). However, the raw image spectrum is predominantly noisy in the same spectral band. Additionally, the active image spectrum also shows a node at $k_y = 4k_0$ which is not detectable in the raw image spectrum. Since the missing Fourier component is within the pass-band of the filter, we must conclude that the node is a feature of the object which was previously buried under the noise. Moreover, a comparison of the SD noise levels between figures \ref{fig:Fig_FFT_Field_Distributions_SiO2}(d) and \ref{fig:Fig_FFT_Field_Distributions_Initial}(d) indicates that they are approximately in the same level. This shows how the selective amplification with the auxiliary source allows us to recover the object features with little noise amplification.

After determining the active image spectrum for each filter-lens integrated system, the final step is a post-processing technique where the noisy portions of the raw image spectrum are substituted with the noise-free portions of the relevant active image spectrum and then deconvolved with an active transfer function. The active image spectrum for the $j^{th}$ spatial filter is first expressed as
\begin{equation}
I^{(j)}_a(k_y) = H_0T_l(k_y)P^{(j)}(k_y)O(k_y),
\label{eq:Active_image}
\end{equation}
where $T_l(k_y)$, $P^{(j)}(k_y)$ are the transfer functions of the silver lens and the $j^{(th)}$ HMM spatial filter, respectively. $O(k_y)$ is the object spectrum and $H_0$ is the incident plane-wave illumination amplitude. The noise-free portion of each active image spectrum is selected by multiplying Eq. \ref{eq:Active_image} with a rectangular function 
\begin{equation}
R^{(j)}(k_y) = \left\{ \begin{array}{cr}
                1 & \hspace{5mm} \bigg |\frac{k_y - k^{(j)}_c}{W^{(j)}}\bigg | \leq \frac{1}{2} \\
                0 & \hspace{5mm} \texttt{otherwise.} \\
                \end{array} \right .
\label{eq:Rect_function}
\end{equation}
of width $W{(j)}$ and centered at $k^{(j)}_c$ defined for the $j^{th}$ integrated system. The equivalent portions of the raw image spectrum are substituted by the selected active image spectra. The resulting total image spectrum can be written as
\begin{equation}
I^{'}(k_y) = I(k_y)\bigg [ 1 - \sum \limits ^{6} _{j=1} R^{(j)}(k_y)  \bigg ] + \sum \limits ^{6} _{j=1}  I^{(j)}_a(k_y)R^{(j)}(k_y),
\label{eq:Total_image}
\end{equation}
where $I(k_y)$ is the raw image spectrum. Similarly, the active transfer function $T_a(k_y)$ is obtained by substituting identical portions the silver lens transfer function with the integrated system transfer function and is expressed as
\begin{equation}
T_a(k_y) = T_l(k_y)\bigg \{ 1 - \sum \limits ^{6} _{j=1} R^{(j)}(k_y)   + \sum \limits ^{6} _{j=1}H_0P^{(j)}(k_y)R^{(j)}(k_y)\bigg \}.
\label{eq:Active_transfer_function}
\end{equation}
The final reconstructed image is obtained by multiplying $I^{'}(k_y)$ from Eq. \ref{eq:Total_image} with the inverse of the active transfer function in Eq. \ref{eq:Active_transfer_function} (i.e., active deconvolution). The values of $W^{(j)}$, $k^{(j)}_c$, and $H_0$ which were used for this reconstruction process are listed in table \ref{tab:Table_Rect} for each filter-lens integrated system. Note that two adjacent rectangle functions should not overlap for Eqs. \ref{eq:Total_image} and \ref{eq:Active_transfer_function} to be valid.
\begin{table}[h]
\small
\centering
\caption{\bf Post-processing parameters}
\begin{tabular}{ccccccc}
\hline
              & $SiO_2$     & $Al_2O3$   & $Si_3N_4$  & $ZrO_2$     & $MoO_3$          & $TiO_2$          \\
\hline
$W^{(j)}$   $(k_0)$  & $0.949$     & $0.476$    & $0.456$    & $0.369$     & $1.145$          & $1.882$         \\
$k^{(j)}_c$ $(k_0)$  & $2.366$     & $3.078$    & $3.544$    & $3.957$     & $4.716$          & $6.228$         \\
$H_0$ $(Am^{-1})$        & $3000$      & $3000$     & $5000$     & $60000$     & $3 \times 10^6$  & $5 \times 10^6$  \\
\hline
\end{tabular}
\label{tab:Table_Rect}
\end{table}

\begin{figure}[htbp]
\centering
\includegraphics[width=\linewidth]{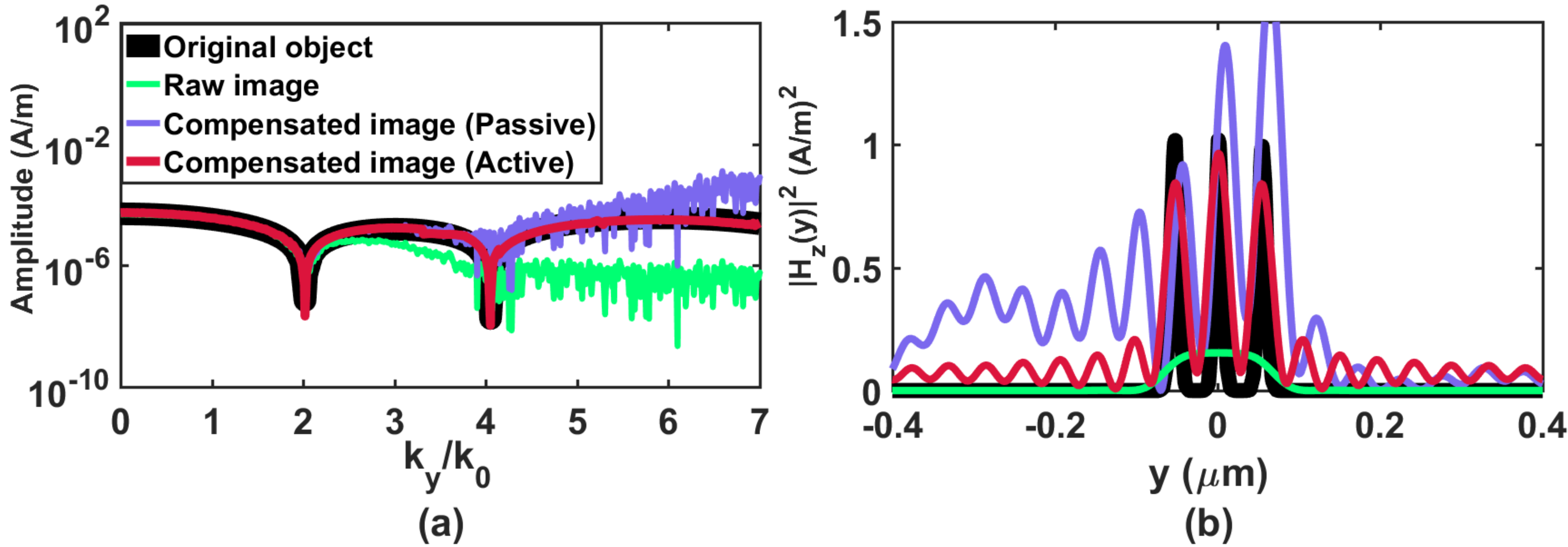}
\caption{(a) Fourier spectra and (b) spatial field distributions of the reconstructed images illustrating the superiority of the active $\Pi$ scheme over passive deconvolution. The imaging system is now capable of resolving objects separated by $\lambda /6$.}
\label{fig:Fig_Field_Distributions_Final}
\end{figure}

The reconstructed image spectrum obtained after the above post-processing steps is shown by the red line in figure \ref{fig:Fig_Field_Distributions_Final}(a). The active compensated spectrum closely follows the object spectrum shown by the black line and is almost noise-free. The raw image spectrum, shown by the green line and previously illustrated in figure \ref{fig:Fig_FFT_Field_Distributions_Initial}(c), is significantly corrupted beyond $k_y \gg 3k_0$ and the passively compensated image spectrum, obtained by a simple deconvolution method \cite{Adams:17,adams2016bringing} shows significant noise amplification. Lastly, figure \ref{fig:Fig_Field_Distributions_Final}(b) compares the amplitude squared of the reconstructed fields with the original object illustrating the significant enhancement of the active $\Pi$ loss compensation scheme. The lens completely fails to resolve the object and passive reconstruction is extremely unreliable due to significant noise amplification whereas, active compensation can resolve the object with a sufficiently high contrast.

In conclusion, we have shown how the $\Pi$ loss compensation scheme when physically implemented with a near-field superlens enhances the ability of the superlens to resolve objects further beyond the diffraction limit. The introduction of a convolved auxiliary source, generated by integrating a HMM spatial filter into the system allows restoration of attenuated components of the image spectrum without amplifying noise. The proof-of-principle presented in this work can be verified in an experimental setting with available plasmonic lenses and hyperlenses. We believe that this work brings us closer to overcoming losses in near-field imaging with MMs and elevates the possibility of realizing a "perfect lens" first envisioned by Pendry. Finally, consistent with this letter, it was brought to our attention that in two other independent works HMM spatial filters integrated with a superlens cavity and a type I HMM were recently used to achieve high-resolution Bessel beam \cite{liu2017nanofocusing} and hyperbolic dark-field lens \cite{shen2017hyperbolic}, respectively.

\section{Funding Information}
Office of Naval Research (award N00014-15-1-2684).

\section{Acknowledgement}

Superior, a high-performance computing infrastructure at Michigan Technological University, was used in obtaining the results presented in this publication.

\end{document}